\documentclass[9pt,twocolumn,twoside]{revtex4-1}

\usepackage{mathtools}
\usepackage{amsfonts}
\usepackage{amssymb}
\usepackage[T1]{fontenc}
\usepackage{dsfont}
\usepackage{array}

\usepackage{braket}

\usepackage{ulem}
\usepackage{xcolor}

\begin{document}
\title{Using excited states and degeneracies to enhance the electric polarizability and first hyperpolarizability}

\author{Ethan L. Crowell}
\author{Mark G. Kuzyk}
\affiliation{Department of Physics and Astronomy, Washington State University, Pullman, Washington  99164-2814}


\begin{abstract}
We investigate the efficacy of boosting the nonlinear-optical response by using novel systems such as those in an excited state or with a degenerate ground state.  By applying the Three Level Ansatz (TLA) and using the Thomas-Reiche-Kuhn (TRK) sum rules as constraints, we find the electric polarizability and first hyperpolarizability of excited state systems to be bounded, but \textit{larger} than those derived for a system in the ground state. It is shown that a system with a degenerate ground state can have divergent polarizabilities and that such divergences are real and not relics of a pathology in the perturbation theory. Furthermore, we demonstrate that these divergences only occur on time scales short compared to the relaxation time of the population difference to an equilibrium value. Such systems provide a way to get ultra-large nonlinear optical response.  We discuss examples of huge enhancements in molecules and double quantum dots using a double quantum well as a model.
\end{abstract}

%

\maketitle

\section{Introduction}
The interaction of a material with an electric field is often well described by the electric polarizability and hyperpolarizabilities. The polarizabilities derive from noting that the electric dipole moment $\mathbf{p} = -e\mathbf{x}$ of a material will depend on the applied electric field $\mathbf{\mathcal{E}}$. Doing a series expansion of the dipole moment in powers of the electric field leads to
\begin{align}
p_i = \mu^{(0)}_i + &\alpha_{ij}\mathcal{E}_j + \beta_{ijk}\mathcal{E}_j\mathcal{E}_k + \dots,
\label{power_series_expansion}
\end{align}
where $\mu^{(0)}_i$ is the static dipole moment, $\alpha$ is the polarizability tensor, and $\beta$ is the first hyperpolarizability tensor \cite{boyd09.01,kuzyk17.01}. Equation \ref{power_series_expansion} is in the frequency domain, but we will treat the static off-resonant limit.  The polarizabilities are readily determined from
\begin{align}
\alpha_{ij} &= \frac{\partial p_i}{\partial\mathcal{E}_j}\bigg|_{\mathcal{E}_j=0}\label{alpha_defn}\\ \intertext{and}
\beta_{ijk} &= \frac {1} {1 + \delta_{j,k}} \cdot \frac{\partial^2 p_i}{\partial \mathcal{E}_j \partial \mathcal{E}_k}\bigg|_{\mathcal{E}_j,\mathcal{E}_k=0}\label{beta_defn},
\end{align}
where $\delta_{j,k}$ is the Kronecker delta.

Large (hyper)polarizabilities are important for a broad range of applications, including telecommunications \cite{chen04.01}, quantum computing \cite{chuan95.01,hutch09.01}, three dimensional nanophotolithography \cite{cumps99.01,kawat01.01}, and the development of materials for cancer therapies \cite{karot04.01,roy03.01}. As such, it is of interest to determine the fundamental characteristics that lead to large polarizabilities.

Such an investigation has been carried out \cite{shafe13.01, kuzyk00.01}, beginning with the closed form expressions for the off-resonant polarizability and first hyperpolarizability,
\begin{align}
\alpha &= 2e^2 \sideset{}{'}\sum_n \frac{x_{0n}x_{n0}}{E_{n0}}\label{sos_alpha_grnd}\\ \intertext{and}
\beta &= 3e^3 \sideset{}{'}\sum_{n,m} \frac{x_{0n}\bar{x}_{nm}x_{m0}}{E_{n0}E_{m0}},\label{sos_beta_grnd}
\end{align}
where $|0\rangle$ is the ground state vector, $E_{nm} = E_n-E_m$, $x_{nm} = \langle n|x|m\rangle$, $\bar{x} = x - x_{00}$, the prime on the sum indicates exclusion of the ground state from the sum, and the system's initial state is assumed to be $\Ket{0}$. The above expressions are readily determined from time-dependent perturbation theory and apply for zero-frequency fields in the absence of damping, though they can be easily generalized to include oscillating fields and relaxation mechanisms \cite{boyd09.01,kuzyk17.01}.

The upper bound on the (hyper)polarizability is determined by applying the three-level ansatz (TLA) and using as constraints the Thomas-Reich-Kuhn (TRK) sum rules given by
\begin{align}
\label{sum_rules}
\sum_n x_{pn}x_{nq}\left(E_n - \frac{1}{2}\left(E_p + E_q\right)\right) = \frac{N\hbar^2}{2m}\delta_{pq},
\end{align}
where $N$ is the number of electrons in the system. The limits obtained are \cite{shafe13.01, kuzyk00.01}
\begin{align}
\alpha_\text{max} &= \frac{N_e e^2\hbar^2}{mE_{10}^2}\label{alpha_lim_grnd}\\ \intertext{and}
\beta_\text{max} &= \sqrt[4]{3}\left(\frac{e\hbar}{\sqrt{m}}\right)^3\left[\frac{N^{3/2}}{E_{10}^{7/2}}\right]\label{beta_lim_grnd}.
\end{align}

The energy difference between the ground and first excited state, $E_{10}$, sets a fundamental limit on the electric polarizability and first hyperpolarizability. These limits have been corroborated by experiment \cite{kuzyk03.02}, potential optimization \cite{zhou07.02, watki09.01, watki11.01, ather12.01, burke13.01}, and calculations on quantum graphs \cite{shafe12.01, lytel13.04, lytel13.01, lytel16.01} though a recent Monte Carlo study utilizing filtered sampling suggests that these limits may be an overestimate by approximately $30\%$ \cite{lytel17.01}.

It is important to reiterate that the above limits are for a system in its ground state. However, by optical pumping a system can be placed in an excited state, or even a linear combination of states. An example of this for two-level systems is a $\theta$-pulse, which can excite the system from the ground state $|0\rangle$ to the linear combination $\cos\theta|0\rangle + \sin\theta|1\rangle$ \cite{steck18.01}. This allows one to reliably occupy the first excited state via a $\pi$-pulse \cite{cao98.01}. Indeed, it is known that the polarizability of a molecule can significantly increase with excitation \cite{roden95.01, vilas18.01, chern05.01, groze01.01, ponde83.01}. A similar trend has been observed for nonlinear optical effects: the hyperpolarizability has been observed to undergo considerable enhancement upon excitation \cite{faust02.01, willi95.01}. Such a situation leads one to ask whether the polarizabilities of such systems--what we call the \textit{excited state polarizabilities}--are bounded in the same sense as above, and whether the excited state polarizabilities can be larger than the aforementioned limits.

In this paper we seek to answer these questions. In section II, we apply the TRK sum rules to derive an upper bound on the excited state polarizabilities and we compare these limits to those of the ground state. This comparison will bring up the role of degenerate states on the polarizabilities, which will be discussed in section III. Finally, we treat an ensemble of degenerate systems and investigate the effects of relaxation processes.

\section{Excited State Limits}
In this section we derive the limits on the polarizability and first hyperpolarizability for a system in the first excited state. We assume the absence of magnetic fields, which makes the hamiltonian invariant under time reversal and allows us to make the simplifying assumption that the dipole moment matrix is real valued.
\subsection{Linear response}
The polarizability in the first excited state can be written as
\begin{align}
\alpha = 2e^2 \sum_{n\neq 1}\frac{|x_{1n}|^2}{E_{n1}}\label{sos_alpha_frst}.
\end{align}
This can be split into positive-definite and negative-definite terms
\begin{align}
\alpha = 2e^2\left(-\frac{|x_{01}|^2}{E_{10}} + \sum_{n > 1}\frac{|x_{1n}|^2}{E_{n1}}\right).
\end{align}

To find a lower bound, we assume the transition moment in the negative-definite term, $x_{10}$, to be at its maximum. This optimal value is defined via the $p=q=0$ sum rule:
\begin{align}
\sum_n |x_{n0}|^2E_{n0} = \frac{N\hbar^2}{2m}\\
\therefore |x_{10}|^2 \leq x_\text{max}^2 = \frac{N\hbar^2}{2mE_{10}}
\end{align}
and requires $x_{0n} = 0,\quad \forall n>1$, which makes $\alpha$ maximally negative. Note that the negative polarizability implies stimulated emission, as one finds for an inverted population as we have here.

Substituting these values of transition moments into Equation \ref{sos_alpha_frst} gives the inequality
\begin{align}
\alpha \geq -\frac{Ne^2\hbar^2}{mE^2_{10}},
\end{align}
which according to Equation \ref{alpha_lim_grnd} is the negative of the limit for a ground state system.

We can likewise determine an upper bound on $\alpha$ by placing all the oscillator strength in the positive terms. Exclusion of the negative terms from the SOS expression gives
\begin{align}\label{eq:alpha-Ineq}
\alpha \leq 2e^2\sum_{n>1} \frac{|x_{1n}|^2}{E_{n1}}.
\end{align}
Recalling the $p=q=1$ sum rule
\begin{align}
\sum_n |x_{1n}|^2E_{n1} = \frac{N\hbar^2}{2m}
\end{align}
with $x_{10}=0$ gives
\begin{align}\label{eq:excStateSumRule}
\sum_{n>1} |x_{1n}|^2E_{n1} = \frac{N\hbar^2}{2m}.
\end{align}
Assuming that the transition between states 2 and 1 dominates -- which gives the largest polarizability, Equation \ref{eq:alpha-Ineq} can be rewritten using Equation \ref{eq:excStateSumRule} to yield
\begin{align}
\alpha &\leq \frac{Ne^2\hbar^2}{mE_{21}^2}.
\end{align}

Thus, the bounds of the polarizability for a system in its first excited state are given by
\begin{align}
-\frac{Ne^2\hbar^2}{mE^2_{10}} \leq \alpha \leq \frac{Ne^2\hbar^2}{mE_{21}^2}.
\end{align}
There are two things of interest to note: first, that the energy difference associated with spontaneous emission from the occupied state $|1\rangle$ is $E_{10}$, and indeed the negative limit for $\alpha$ scales with $E_{10}$. Likewise, the smallest energy difference associated with absorption is $E_{21}$, and indeed the positive limit for $\alpha$ scales with $E_{21}$. Second, for a degenerate first excited state with $E_2 = E_1$, the limit diverges. This intriguing possibility for making arbitrarily-large polarizability is discussed later.

\subsection{Quadratic response}
The limit of the first hyperpolarizability from an excited state parallels the original work \cite{kuzyk00.01,kuzyk06.03}. The first hyperpolarizability of the first excited state is given by
\begin{align}
\beta &= 3e^3 \sideset{}{'}\sum_{n,m} \frac{x_{1n}\bar{x}_{nm}x_{m1}}{E_{n1}E_{m1}},\label{sos_beta_frst}
\end{align}
where we redefine the barring operator as $\bar{x} = x - x_{11}$ and the prime now indicates exclusion of the first excited state from the sum. We can use the $p=n, q=1$ sum rule to get the excited state variant of the dipole-free SOS expression for $\beta$ \cite{kuzyk05.02}:
\begin{align}
\beta = -3e^3\sideset{}{'}\sum_{n\neq m} x_{1n}x_{nm}x_{m1}\left(\frac{1}{E_{n1}E_{m1}} - \frac{2E_{n1} - E_{m1}}{E_{m1}^3}\right)\label{beta_df}
\end{align}
where, again, the prime on the sum indicates exclusion of the first excited state. Applying the TLA gives us
\begin{align}
\beta = -3e^3 x_{10}x_{02}x_{21}&\left(-\frac{2}{E_{21}E_{10}}  + \frac{2E_{21} + E_{10}}{E_{10}^3}\right.\nonumber\\
& + \left.\frac{2E_{10} + E_{21}}{E_{21}^3}\right)
\end{align}

Using the $p=q=0$ and the $p=q=1$ sum rules, the upper bound on the transition elements is \cite{kuzyk06.03}
\begin{align}
|x_{01}x_{12}x_{20}| &\leq \frac{E}{\sqrt{1-E}}x_{10}^3X\sqrt{1-X^4}\nonumber\\
&\equiv \frac{E}{\sqrt{1-E}}x_{10}^3\sqrt[4]{\frac{1}{3}}\sqrt{\frac{2}{3}}G(X),\label{transition_moments_bound}
\end{align}
where the function
\begin{align}
G(X) = \sqrt[4]{3}\sqrt{\frac{3}{2}}X\sqrt{1-X^4}
\end{align}
has as a maximum value $G(\sqrt[4]{3}) = 1$. Furthermore, from the $p=q=0$ sum rule we have
\begin{align}
|x_{10}|^3 \leq \left(\frac{N\hbar^2}{2mE_{10}}\right)^{3/2} .
\end{align}

Putting this all together results in the inequality
\begin{align}
|\beta| \leq \frac{1}{2}\sqrt[4]{3}&\left(\frac{e\hbar}{\sqrt{m}}\right)^3N^{3/2}\nonumber\\
&\times \left(\frac{1}{E_{10}} + \frac{1}{E_{21}}\right)^{7/2}f(E),
\end{align}
where we have defined $E = \frac{E_{10}}{E_{20}} \leq 1$ and
\begin{align}
f(E) = -2 E^2&(1-E)^2 + 2E^4 + (1-E)E^3\nonumber\\
 &+ 2(1-E)^4 + E(1-E)^3.\label{f(E)_function}
\end{align}
Finally, it is easy to show that $f(E)$ is bounded above by 2 and below by $\frac{1}{4}$, giving us a fundamental limit on the excited state polarizability:
\begin{align}\label{eq:exciteStateMax_alt}
\beta_\text{max}^\text{exc. st.} = \sqrt[4]{3}\left(\frac{e\hbar}{\sqrt{m}}\right)^3N^{3/2}\left(\frac{1}{E_{10}}+\frac{1}{E_{21}}\right)^{7/2}.
\end{align}
which can be rewritten as
\begin{align}\label{eq:exciteStateMax}
\beta_\text{max}^\text{exc. st.} = \sqrt[4]{3}\left(\frac{e\hbar}{\sqrt{m}}\right)^3\frac{N^{3/2}}{E_{10}^{7/2}} \left(\frac{E_{20}}{E_{21}}\right)^{7/2}.
\end{align}

The above limits were derived using the three level ansatz with states $|0\rangle$, $|1\rangle$ and $|2\rangle$. One could alternatively use the states $|1\rangle$, $|2\rangle$, and $|3\rangle$; the former might miss an important transition form a higher state while the latter misses the stimulated emission process.  The appendix treats the latter case and also determines the hyperpolarizability in the four-level ansatz.  The alternative three-level ansatz that omits the ground state gives results identical to those for ground state systems with $E_{10}\rightarrow E_{21}$.  The four-level ansatz gives $\beta_\text{max}^\text{4L} = 0.957 \beta_\text{max}^\text{3L}$, which is thus within a few percent of the three-level ansatz limit $\beta_\text{max}^\text{3L}$.  As such, using sates $|0\rangle$, $|1\rangle$ and $|2\rangle$ is appropriate.

Note that the above calculations only hold for nonrelativistic systems. It has been predicted that the fundamental ground state limits decrease with first order relativistic corrections \cite{dawson15.01}. This is largely due to the relativistic transition moments being smaller than their nonrelativistic counterparts. What effects this might have on the excited state limits will be the subject of future considerations.

\subsection{Comparison to ground state limits}
The excited state limits are compared with those derived for the ground state by normalizing them by the ground state limits to obtain the \textit{intrinsic} excited state limits, yielding
\begin{align}\label{eq:intLimitAlpha}
-1 \leq &\alpha_\text{int}^\text{exc. st.} \leq \left(\frac{E_{10}}{E_{21}}\right)^2\\ \label{eq:intLimitBeta}
-\left(\frac{E_{20}}{E_{21}}\right)^{7/2}\leq &\beta_\text{int}^\text{exc. st.} \leq \left(\frac{E_{20}}{E_{21}}\right)^{7/2}.
\end{align}
Since $E_{21}\leq E_{20}$, the excited state hyperpolarizability limit is always at least as large as the ground state limit. The intrinsic excited state polarizability can also exceed unity, though this is only when the first excited state energy is more proximate to the second excited state energy than it is to the ground state energy.

The physics behind an intrinsic (hyper)polarizability exceeding unity can be understood from the point of view of scaling. One typically understands the normalization of the polarizabilities by the limits as a way of introducing scale-invariant quantities: if we scale the system by $\epsilon$, i.e. $x\rightarrow \epsilon x$, then the energies scale as $E_n \rightarrow \epsilon^2 E_n$ and the transition moments scale as $x_{nm} \rightarrow x_{nm}/\epsilon$ \cite{kuzyk13.01}. Thus, we have $\alpha \rightarrow \alpha/\epsilon^4$ and $\beta \rightarrow \beta/\epsilon^7$. Furthermore, by the same rules we see that the intrinsic polarizabilities scale in the same way. Thus, the intrinsic polarizabilities are invariant as the size of the system changes. The fact that the excited state intrinsic polarizabilities can exceed unity then suggests that as we increase the proximity of the energy levels, the excited state wavefunctions expand faster than the ground state wavefunctions, thus causing the polarizabilities to increase more quickly.

It is also interesting to consider the situations where the excited state limit is smaller than the ground state limit. Note that $\alpha_\text{int}^\text{exc. st.}<1$ when $E_{10}<E_{21}$. Further, recall that $E_{10}$ can be associated with spontaneous emission from the first excited state to the ground state. When $E_{10}<E_{21}$, the spontaneous emission process, which has a negative definite contribution to $\alpha$, has a larger contribution to the polarizability than do the virtual absorption processes. That is, the system is more likely to undergo emission than absorption, and so the optimal value decreases.

\subsection{Monte carlo}
Although the upper bound on the intrinsic values can be larger than unity, this does not imply that intrinsic values of actual systems will be. To determine the possibility of intrinsic values greater than one, we implement a statistical approach in which we randomly generate sets of eigenenergies and transition moments. The moments and energies are chosen such that the diagonal sum rules are obeyed exactly, while the off-diagonal sum rules are obeyed to within a standard error $\epsilon$. The technique for choosing such sets has been described elsewhere \cite{lytel17.01}.

\begin{figure}
\centering
\includegraphics[width=\columnwidth]{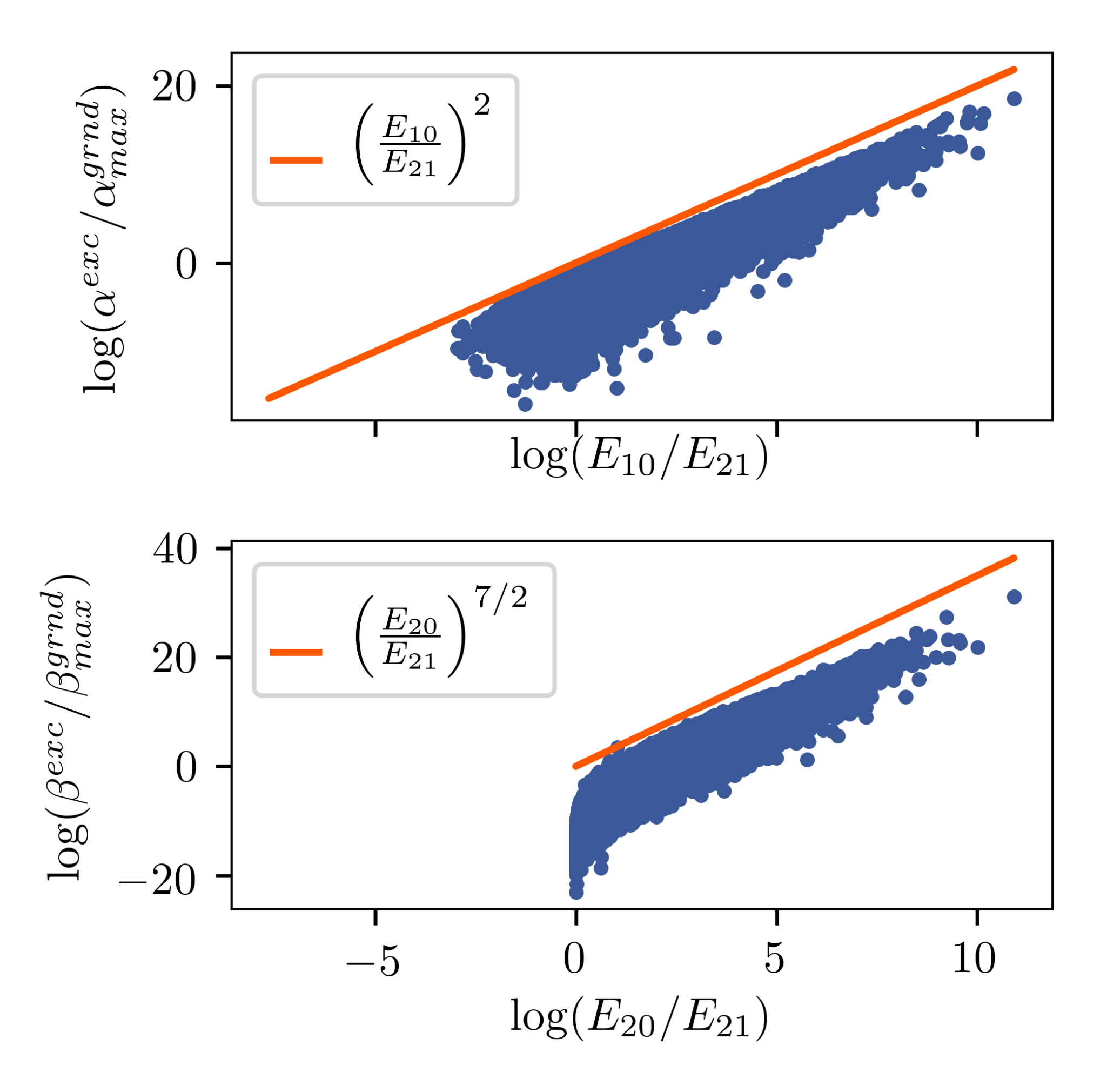}
\caption{Intrinsic excited state polarizabilities as function of energy parameter. The orange line indicates the limits derived in the previous section and the blue points are the values obtained from randomly generating energies and transition moments. One can see that the limits are well obeyed.}
\label{fig_monte_carlo_energy_correlation}
\end{figure}

Fig \ref{fig_monte_carlo_energy_correlation} plots the intrinsic excited state limits for Monte-Carlo-generated ``molecules" (points) and the intrinsic limits given by Equations  \ref{eq:intLimitAlpha} and \ref{eq:intLimitBeta} (line) as a function of the energy ratios. The excited state intrinsic polarizabilities and hyperpolarizabilities are much larger than unity, and larger values of $E_{10}/E_{21}$ tend to give larger intrinsic values, in agreement with Equations  \ref{eq:intLimitAlpha} and \ref{eq:intLimitBeta}.  Also, all points are seen to fall below the limit line, so the calculated limits are obeyed.

As the energy ratio $E_{20}/E_{21}$ becomes large, the points fall lower below the calculated limits, implying that they overestimate the response when the excited states become degenerate.  Such behavior is familiar when the the sum rules are truncated, as is done in applying the TLA.  This can lead to unphysical behavior\cite{kuzyk15.01} and give rise to a gap between the limit and physical systems \cite{lytel17.01}. Whether this is the case here, or if this apparent gap is just the result of under-sampling due to limited computation time is currently unknown and will be the subject of future studies.

\begin{figure}
\centering
\includegraphics[width=\columnwidth]{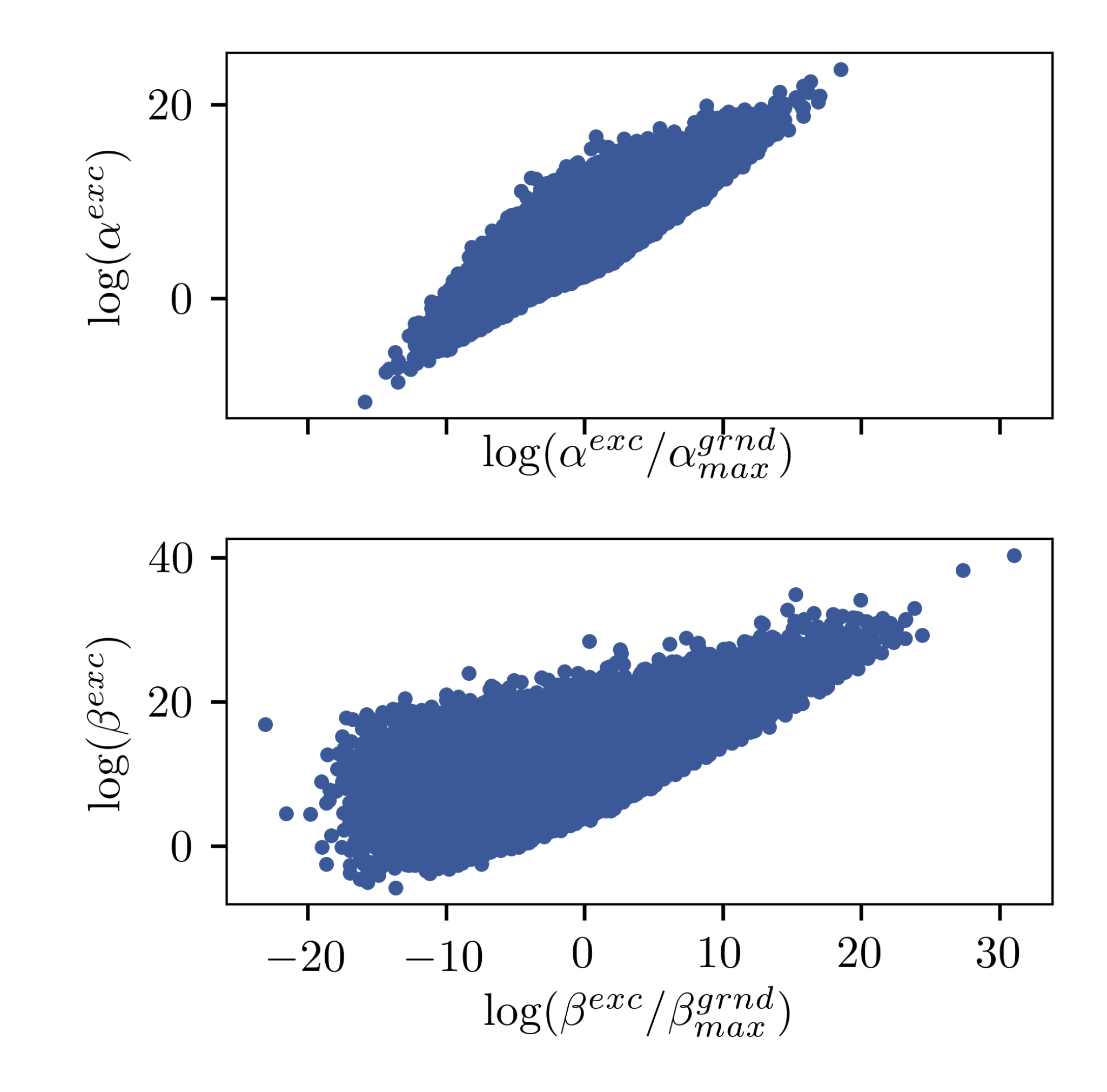}
\caption{Unscaled vs intrinsic polarizabilities. The fact that the polarizabilities can have large scaled and unscaled values assures that we are not simply minimizing the maximum ground state response when seeing large intrinsic values.}
\label{fig_monte_carlo_raw_vs_int}
\end{figure}

We stress that intrinsic quantities are not the sole focus for optimization, but rather are an intermediate step in finding materials that scale in a way that optimizes the raw values.  A pitfall to the approach of optimizing the intrinsic polarizabilities is that, although large intrinsic values can result from large excited state polarizabilities, they can also result from small polarizabilities provided sufficiently small ground state limits.  However, this appears not to be the case with the excited state polarizabilities as illustrated in Fig \ref{fig_monte_carlo_raw_vs_int}; large intrinsic values can also correspond to large unscaled values that are enhanced many orders of magnitude over what is possible in the ground state.

An interesting example of large polarizability of an excited-state system is the Rydberg atom, where one of the valence electrons is excited to a large principle quantum number. The atoms are surprisingly stable due to the negligible overlap of the excited states with the lower lying states \cite{eichm13.01}. It is known that Rydberg atoms tend to have very large polarizabilities, varying as $n^7$, where $n$ is the principle quantum number \cite{hecko17.01}. Because $n$ is quite large for Rydberg atoms (upwards of 30), the polarizabilities can reach values on the order of $10^{12}$ $a_0^3$ \cite{yerok16.01, osull85.01, osull86.01}.

To illustrate the enhancement experienced by Rydberg atoms, we consider Rubidium. In the ground and first excited states, the valence electron occupies the $^2S_{1/2}$ and $^2P_{1/2}$ orbitals, respectively, with energies given in Table \ref{Rb_energy_levels}. Hence, the ground state limit for Rubidium is
\begin{align}
\alpha_\text{max} & = \frac{e^2\hbar^2}{mE_{10}^2}\nonumber\\
&=2.25688\times 10^8 a_0^3,
\end{align}
where we have assumed, for simplicity, that only the excited valence electron interacts with the field. The polarizability for Rubidium has been measured and calculated for excited states up to $n=50$, with some characteristic values shown in Table \ref{Rb_energy_levels}. Thus, the intrinsic polarizability for Rubidium $50^2P_{1/2}$ is $5.27\times 10^3$ $a_0^3$, much larger than the limit of unity for ground state systems. Although the states pertinent to the dynamics of a Rydberg atom's optical response are not the ground, first, and second excited states treated above, we still see the same trend: by exciting to a higher energy state where the energy levels are more dense, one can exceed the fundamental limits placed on the ground state polarizabilities.

We can compute the excited state limit if we make the associations $E_1 \rightarrow E_n=50 ^2P_{3/2}$, $E_2 \rightarrow E_{n-1}=49 ^2S_{1/2}$, and $E_0 \rightarrow E_{n+1}=48^2S_{1/2}$. Note that $48^2S_{1/2}$ and $49^2S_{1/2}$ are associated with $E_0$ and $E_2$ because they are the closest adjacent energy levels to $50 ^P_{3/2}$ with a nonzero electric dipole transition moment with it. The excited state limit for $50^2P_{3/2}$ is then seen to be $1.89\times 10^{18}$ $a_0^3$, which is still well above the reported value.

Based on the analysis presented in this section, it is highly likely that molecules exist with an unremarkable hyperpolarizability, but when excited may have an large hyperpolarizability if that excited state is nearly degenerate.

\begin{table}[htbp]
\centering
\begin{tabular}{cccc}
\hline
Level& Energy ($E_h$) & Polarizability ($a_0^3)$ & Ref.\\
\hline
$^2S_{1/2}$ & 0.000 &  $3.20\times 10^2$ & \cite{sanso06.01, grego15.01}\\
$^2P_{1/2}$ & $0.0000665649$ &   & \cite{sanso06.01}\\
$50^2S_{1/2}$ & $0.000178040$ & $2.03\times 10^{11}$ & \cite{sanso06.01, yerok16.01}\\
$48^2S_{1/2}$& $0.000178019$ &  & \cite{sanso06.01}\\
$50^2P_{1/2}$ & $0.000178026$ & $1.19\times 10^{12}$ & \cite{sanso06.01, yerok16.01}\\
$49^2S_{1/2}$ & $0.000178028$ &  & \cite{sanso06.01}\\
\hline
\end{tabular}
\caption{Energy levels and static polarizabilities for Rubidium.}
\label{Rb_energy_levels}
\end{table}

\subsection{Comparison to multi-electron systems}
Here we consider multiple noninteracting electrons in an analysis similar to that of Burke and Atherton.\cite{ather16.01} The fact that pre-excitation can enhance the intrinsic polarizability has interesting consequences on such multi-electron systems. To illustrate, consider a single-particle energy spectrum associated with an external potential. First, suppose there is a single electron in the first excited state, and compare this to the case where there are three electrons in the ground state configuration, where the ground state is doubly occupied and a single electron occupies the first excited state.  For the moment, assume that only the electron at the Fermi level interacts with the field. The polarizability $\alpha$ for the two systems are given by
\begin{align}
\alpha_\text{multi} &= 2e^2\sideset{}{'}\sum_{n>1}\frac{|x_{n1}|^2}{E_{n1}}\\
\alpha_\text{single} &= 2e^2\sideset{}{'}\sum_{n>1}\frac{|x_{n1}|^2}{E_{n1}} - 2e^2\frac{|x_{10}|^2}{E_{10}}
\end{align}
where, in the multi-electron case, it is understood that $x = x_1 + x_2 + x_3$. We see immediately that the polarizability for the multi-electron system is larger. This is because the lower states are not available to the upper electron for transition due to Pauli exclusion and thus the negative terms in $\alpha$ are absent. Furthermore, recall that to find the upper limit of the excited state system, we neglected the lower states, which corresponds precisely to the multi-electron system. Thus, both systems have the same upper limit, which scales as $1/E_{21}^2$; however, the single excited electron will likely have a smaller intrinsic polarizability due to the negative contribution from the stimulated emission term.

The hyperpolarizability behaves slightly differently. For the multi-electron case where only the fermi level electron can interact with the field, we have
\begin{align}
\beta_\text{max} = \sqrt[4]{3}\left(\frac{e\hbar}{\sqrt{m}}\right)^3\frac{1}{E_{21}^{7/2}}
\end{align}
However, for the single excited electron we have
\begin{align}
\beta^\text{single}_\text{max} = \sqrt[4]{3}\left(\frac{e\hbar}{\sqrt{m}}\right)^3\frac{1}{E_{21}^{7/2}}\left(\frac{E_{20}}{E_{10}}\right)^{7/2}
\end{align}
We see that the single excited electron may have a \textit{larger} excited nonlinear response than the multi-electron case.  This is likely due to the fact that the terms in SOS expression for $\beta$ with transitions to the ground state may have a positive contribution due to the numerator not being positive definite as is the case for the polarizability.

If all of the electrons in the multi-electron system can be excited, the polarizability is of the form
\begin{align}
\alpha^\text{multi} &= 2e^2\sum_{n>1}\frac{|x_{n1}|^2}{E_{n1}} + 4e^2\sum_{n>1}\frac{|x_{n0}|^2}{E_{n0}} + 2e^2\frac{|x_{01}|^2}{E_{10}}
\end{align}
The hyperpolarizability can be broken into transitions of two different types. In Type I transitions, an electron is excited to a higher state, undergoes a transition between excited states, then de-excites to the starting state. The SOS term for this contribution is
\begin{align}
\beta_\text{I}^\text{multi} &= 3e^3\sum_{n,m>1}\frac{x_{1n}\bar{x}_{nm}x_{m1}}{E_{n1}E_{m1}} + 6e^3\sum_{n,m>1}\frac{x_{0n}\bar{x}_{nm}x_{m0}}{E_{n0}E_{m0}}\nonumber\\
&+ 3e^3\sum_{n>1}\left(\frac{x_{0n}\bar{x}_{n1}x_{10}}{E_{n1}E_{10}}+ \text{c.c.}\right)
\end{align}
A Type II transition involves the fermi level electron being excited, one of the lower level electrons populating the just vacated first excited state, then the initial electron de-exciting into the recently vacated ground state. The SOS term for this process can be written as
\begin{align}
\beta_\text{II}^\text{multi} &= -6e^3\sum_{n>1}\frac{x_{1n}\bar{x}_{01}x_{n1}}{E_{n1}E_{10}} .
\end{align}
It is difficult to say more than this without knowing the energy spectrum and the geometry of the wavefunctions. A study on the optimal many-electron nonlinear response can be found in \cite{ather16.01}.  There is no reason to doubt that the excited state response can be enhanced in the excited state over the ground state configuration for particular systems given that the calculations of the excited state limits as quantified by Equation \ref{eq:exciteStateMax} applies to the many-electron case and works even when there are interactions between the electrons.

\section{Role of degenerate states}
In the previous section we saw that a near degeneracy can enhance the polarizabilities. To investigate this enhancement, we compute the polarizability using finite fields and compare to the perturbative treatment for a double-well potential. That analysis provides insights into the physics of the divergence. Finally, we will investigate the effects of interactions and the relaxation mechanisms. For simplicity we will focus our discussion entirely on the role of degenerate states and thus assume the system to be in a ground state that's quasi-degenerate with the first excited state.

\subsection{Finite fields}
We numerically simulate the optical response of an electron in an infinite well of width $L$ with a rectangular barrier in the middle of variable height but with fixed width $d$.  The time independent Schr{\"o}dinger equation for the system is, in atomic units,
\begin{align}
-\frac{1}{2}\nabla^2 \psi(x) + V(x)\psi(x) = E\psi(x),
\end{align}
where the potential is
\begin{align}
V(x) =
\begin{cases}
0, &0 < x < \frac{L-d}{2}\\
V_0, &\frac{L-d}{2} < x < \frac{L+d}{2}\\
0, &\frac{L+d}{2} < x < L\\
\infty, &\text{otherwise}.
\end{cases}
\end{align}\

\begin{figure}
\centering
\includegraphics[width=\columnwidth]{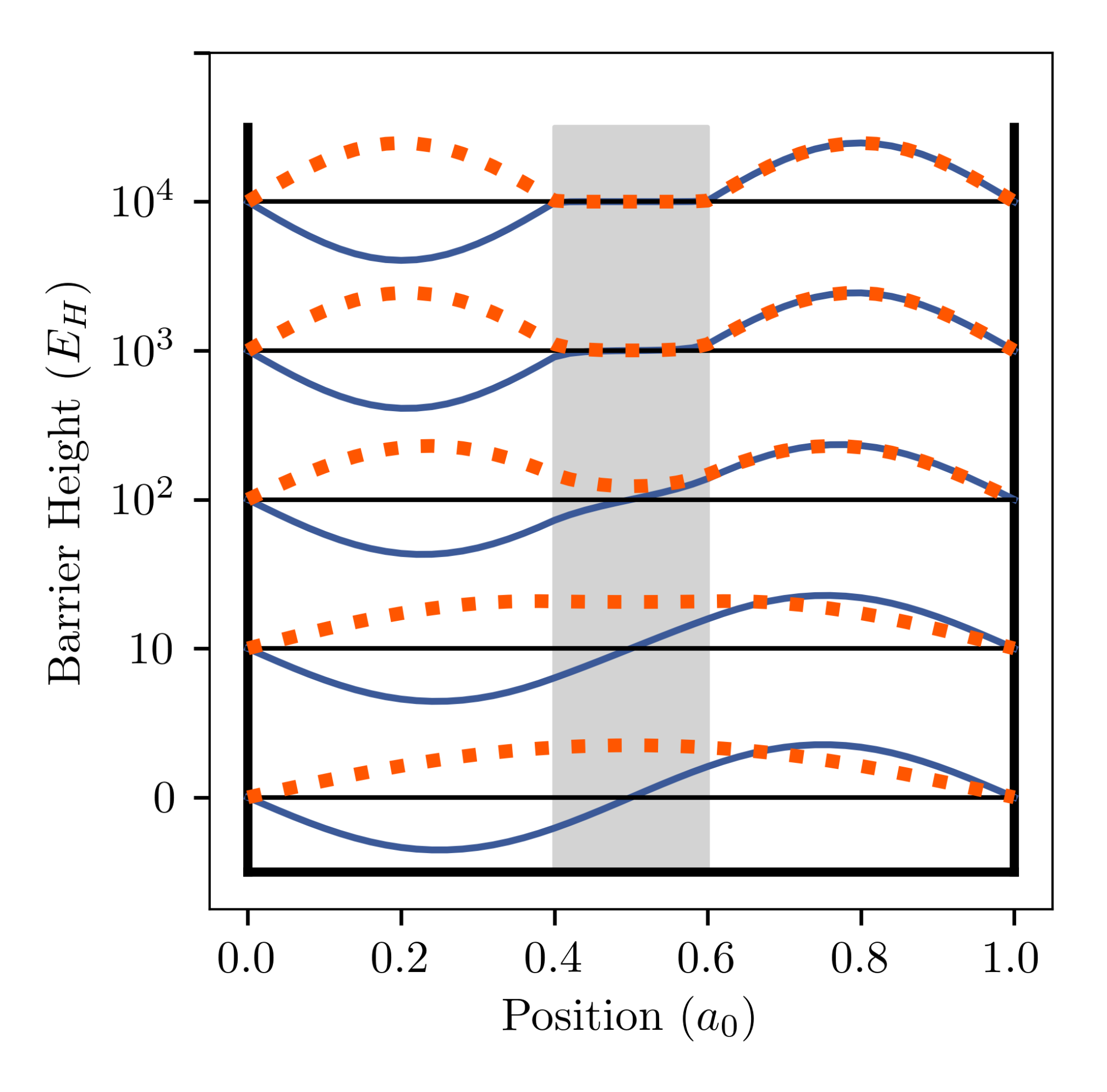}
\caption{Two lowest energy eigenfunctions as a function of barrier height. As the barrier height increases, the eigenfunctions inside the barrier become suppressed and the states approach degeneracy.}
\label{fig_increasing_height}
\end{figure}

The eigenenergies and eigenfunctions were determined using a standard finite difference scheme: position space was represented by a discrete lattice $x = \left[x_0, x_1,\dots, x_N\right]$ which results in a discrete representation of the wavefunction $\psi = \left[\psi(x_0), \psi(x_1),\dots,\psi(x_N)\right]$. Then, using a finite difference approximation for the second derivative, the Sch{\"o}dinger equation can be written as the coupled system of equations
\begin{align}
-\frac{\hbar^2}{2m}\frac{\psi(x_{n+1}) - 2\psi(x_n) + \psi(x_{n-1})}{dx^2} + V(x_n)\psi(x_n) = E\psi(x_n).
\end{align}
This system is then solved using a standard linear algebra package (e.g. numpy.linalg in Python)
Fig \ref{fig_increasing_height} shows the eigenfunctions for various barrier heights. As the barrier height increases, the wavefunction inside the barrier becomes suppressed, resulting in the ground and first excited state curvatures on both sides to become identical.  Hence, the two states become degenerate as the barrier height approaches infinity.

\begin{figure}
\centering
\includegraphics[width=0.5\textwidth]{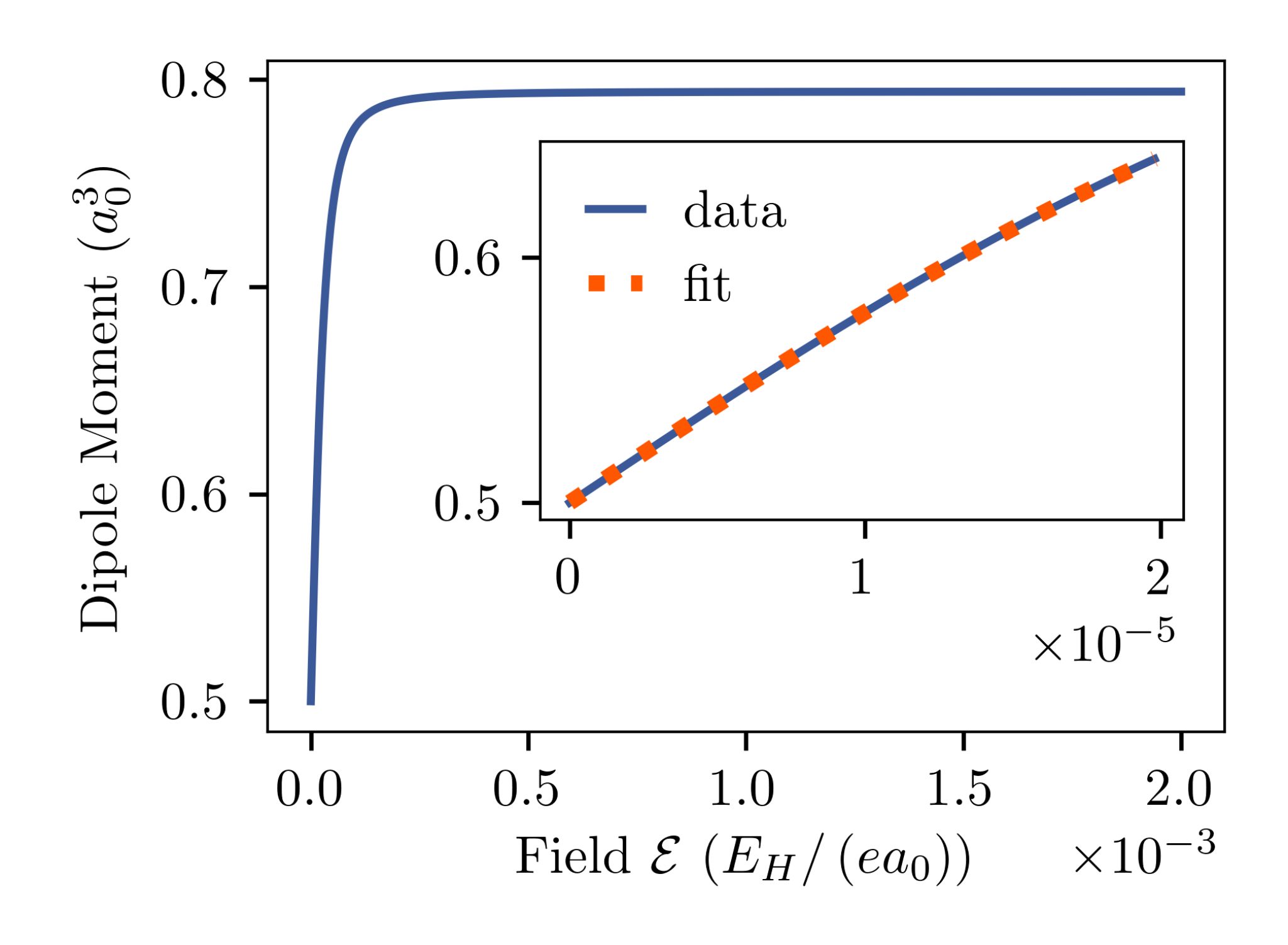}
\caption{Induced electric dipole moment as a function of electric field strength. In the inset, the orange dotted line shows the polynomial fit.}
\label{fig_finite_fields}
\end{figure}

The first step in the finite fields method is to compute the induced dipole moment $p$ for a large number of electric field strengths $\mathcal{E}$, resulting in a numerical representation of the function $p(\mathcal{E})$. An example of this function for a barrier height of 2000 $E_H$ can be seen in Fig \ref{fig_finite_fields}. Note that as the field strength increases, the induced dipole moment appears to saturate at 0.8 $a_0^3$, which corresponds to the particle being localized in the right well. At infinite field strength the dipole moment saturates at 1 $a_0^3$ due to the particle's inability to penetrate the infinite walls on either side of the box.

\begin{figure}
\centering
\includegraphics[width=0.5\textwidth]{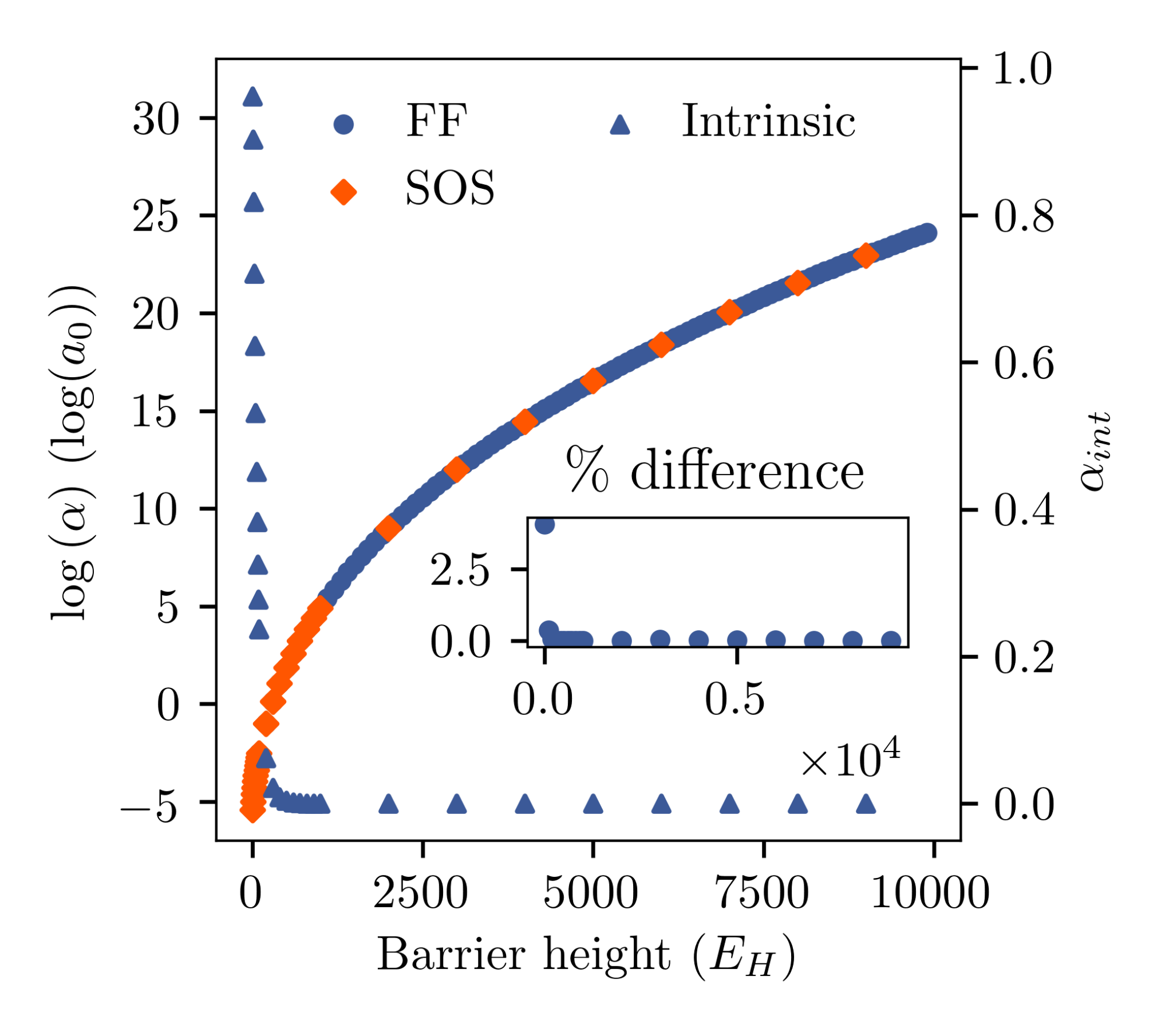}
\caption{Polarizability as a function of barrier height. The polarizability is seen to diverge as the barrier height increases and the system approaches degeneracy, while the intrinsic polarizability approaches zero. The divergence is present in both the perturbative and finite fields calculations, suggesting that it is a real effect. The inset shows the percent difference between the finite fields and perturbative treatments.}
\label{fig_finite_fields_vs_sos}
\end{figure}

With the function $p(\mathcal{E})$ in hand, one can fit it with a polynomial (orange dotted line in the figure) and determine the polarizabilities by comparison with Equation \ref{power_series_expansion}. This procedure is repeated for several barrier heights. For comparison, we use  the states for zero field to compute the SOS expression for $\alpha$, Eq \ref{sos_alpha_grnd}. The results are shown in Fig \ref{fig_finite_fields_vs_sos}.

\begin{figure}
\centering
\includegraphics[width=0.5\textwidth]{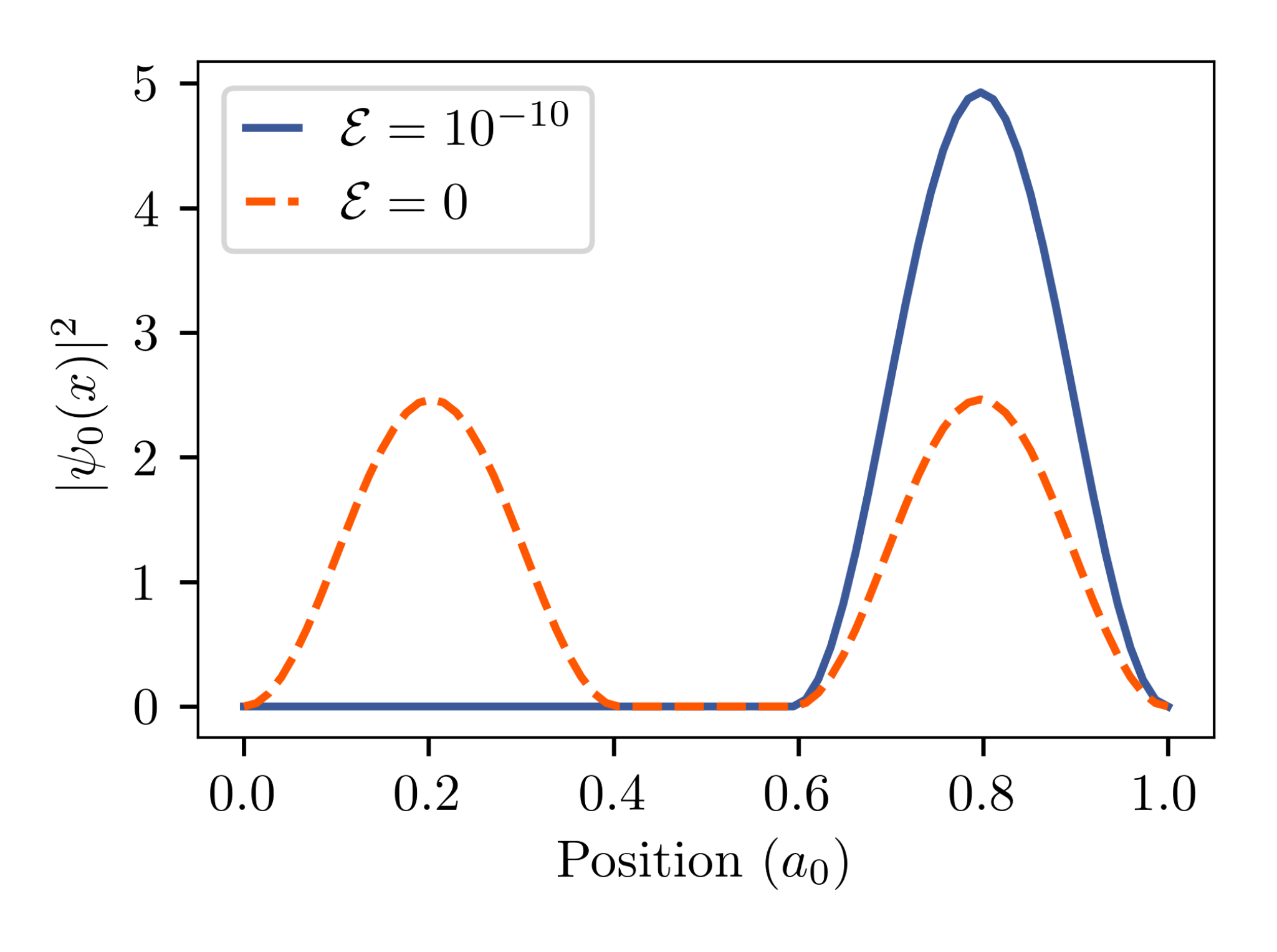}
\caption{Ground state wavefunction before and after application of an infinitesimal field. Note the discontinuous jump of the mean position of the electron from 0.5 $a_0$ to 0.8 $a_0$, which corresponds to a divergent polarizability. The data was taken for a field strength of $\mathcal{E} = 1\times 10^{-10}$ $E_H / (ea_0)$.}
\label{fig_phys_origin}
\end{figure}

There is excellent agreement between the finite fields method and the perturbation method, and both methods diverge as the system approaches degeneracy. This suggests that these large polarizabilities are real and not just the result of perturbation theory breaking down. Indeed, the physical origin of the divergence can be readily elucidated. Fig \ref{fig_phys_origin} shows the ground state wavefunction for zero field -- where we have assumed the electron to be delocalized -- and for an infinitesimal field, where the ground state is unique. It is seen that the electron immediately localizes upon application of the field. This corresponds to a nearly discontinuous change in the mean position of the electron which, in light of Eq \ref{alpha_defn}, implies a truly divergent polarizability. Note that to do an analagous numerical study for the first hyperpolarizability, one must break the centro-symmetry of the system. Furthermore, the barrier heights used in the above calculation (up to $10^4$ E\textsubscript{h} $\sim 10^5$ eV) are much larger than those present in physical systems with similar architecture, e.g. double quantum dots ($\sim 10^2$ meV). However, the preceding work with the symmetric well establishes that, in principle, states can be degenerate \textit{and} have non-zero transition moment between them, which is key to having real divergences in the optical response. The specific case of numerical studies of the effect of degeneracies on the nonlinear response will be relegated to future work.  Here we have shown analytically that this should be the case.

It is important to note that the divergence is not present at degeneracy. For infinite barrier height, the delocalized state cannot tunnel through the barrier, preventing the electron from occupying the ground state of the perturbed system. Indeed, the time it takes the electron to tunnel through the barrier is inversely proportional to the energy difference between the two states. The energy difference between the states is unchanged to first order in perturbation theory, whence we find the time to tunnel $\tau$ is
\begin{align}
\tau = \frac{1}{E_{10}}.
\end{align}
Thus, the response time also diverges. Indeed, the intrinsic polarizability $\alpha / \alpha_{\text{max}}$ goes to zero as the barrier height becomes infinite, and thus we can conclude that the response time diverges more quickly than the plarizability. If we recall that the figure of merit (FOM) for the electric polarizability can be defined as the polarizability divided by the response time, we see that although the polarizability diverges, the FOM remains finite.

The above results raise the question of whether a real system can be made to exhibit this near divergence. Systems with near-degenerate ground states can be found in nature, with examples ranging from cyanine dyes \cite{tolbe93.01} to Bose-Einstein-Condensates in a double well \cite{milbu97.01,trenk16.01} to double quantum dots that can be tuned to degeneracy \cite{petta15.01, hayas03.01}. However, it should be stressed that it is not sufficient for the system to have a degenerate ground state for the enhancement to be present. For example, if there is no transition element between the two degenerate states, then the enhancement vanishes. Thus, the system must be in an appropriate linear combination of states such that they are degenerate and there is nonzero transition strength between them. The aforementioned $\theta$-pulse may be a possibility in populating such linear combinations.

Populating specific electronic states in real molecular systems can be challenging given the presence of molecular vibrations. It is well known that upon an electronic excitation, the nuclei of a molecule will adjust to new equilibrium positions via molecular vibrations (the Franck-Condon effect). This adjustment results in a red shift of the resonant frequency \cite{cao98.01}. However, this issue has been successfully circumvented, at least in theory, by using "chirped" pulses; that is, a pulse with a frequency increasing in time \cite{cao98.01,cao00.01}. With such a pulse, the light is initially resonant with the excited state, and thus excites the electron. However, because the resonant frequency gets redshifted with time and the optical frequency is increasing, the optical field cannot de-excite the electron, thus leaving the excited state populated.  Similarly, if the chirp is drawn out, then the optical frequency can be treated as an adiabatic variable, thus allowing one to tune the electronic energies. These methods have been predicted to successfully achieve up to 99\% population inversion, and could conceivably be used to create linear combinations by adjusting the pulse duration and/or pulse intensity.

Alternatively, one may be able to take advantage of the difference in time scales between molecular vibrations and electronic excitations. The typical time-scale for molecular vibrations is $10^{-13}$ s, while the typical time scale for electronic transitions is $10^{-15}$ s. Furthermore, the theta-pulses mentioned in the introduction are on the order of $10^{-15}$ s \cite{cao98.01}, which makes it feasible to create the appropriate states and perform the measurement before the nuclei can appreciably move.

A perhaps more simple avenue are "artificial molecules" made up of the aforementioned coupled quantum dots. Such devices allow experimentalists exceptional control over the environment of the "molecule" as well as the strength of coupling between them \cite{petta15.01, hayas03.01} which may help circumvent the complications of real molecular systems.


\subsection{Effects of damping}
The above calculations only treat a single electron, isolated from any outside interactions. We would like to include relaxation mechanisms into the dynamics to determine whether the aforementioned divergences are physically feasible. As such, it is pertinent to investigate the behavior of an \textit{ensemble} of such near degenerate systems. Before doing so, however, we will first treat a single electron in the presence of damping.

\subsubsection{Single electron with damping}
\begin{figure}
\centering
\includegraphics[width=\columnwidth]{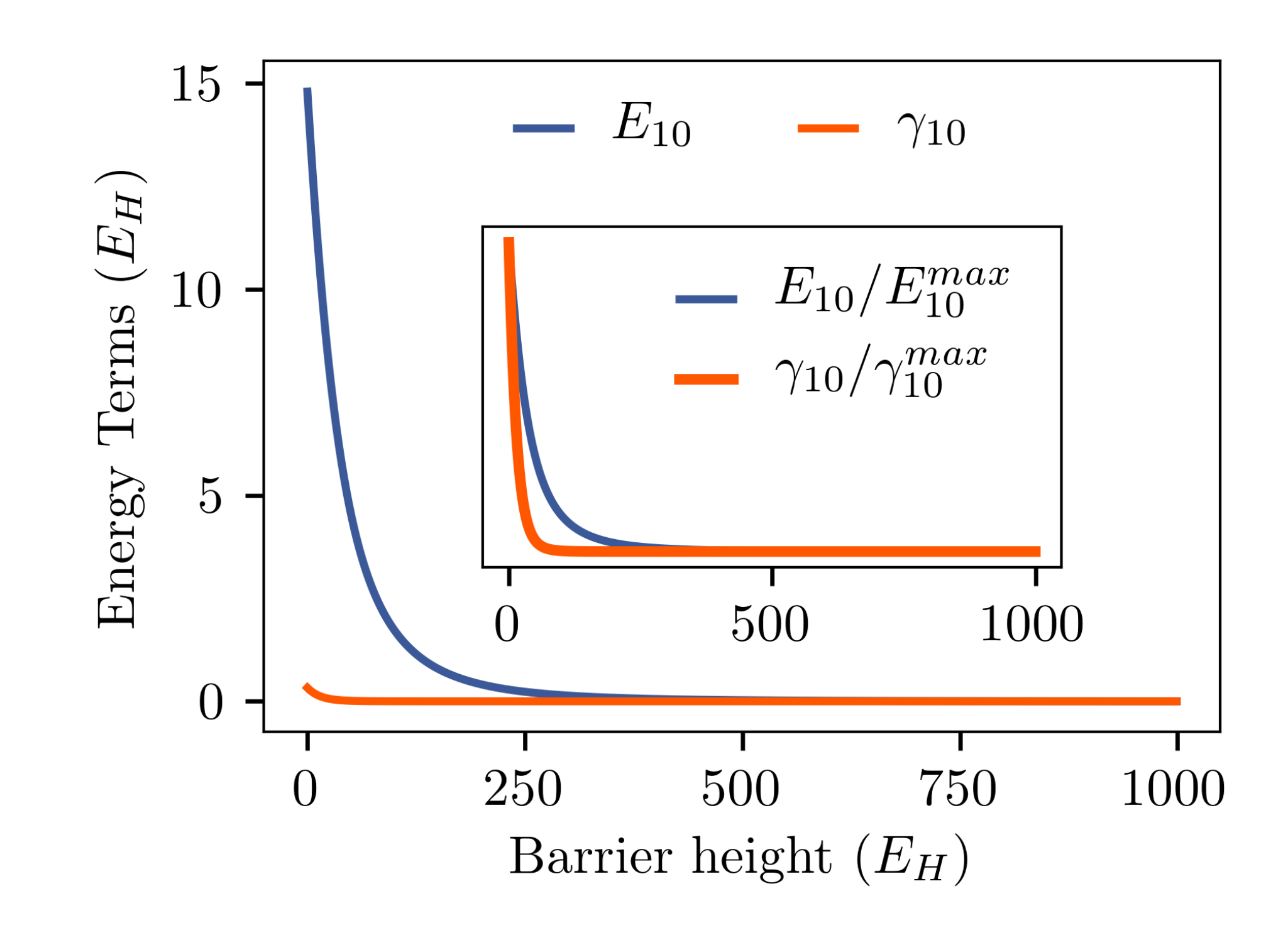}
\caption{The pertinent energy difference $E_{10}$ along with the corresponding damping term $\gamma_{10}$.  The damping term approaches zero in the degenerate limit more quickly than does the energy difference, thereby not supressing the divergence of the polarizabilities.}
\label{damping_effects_alpha}
\end{figure}

To include damping into the single electron scheme, one effectively lets $E_{n0}\rightarrow E_{n0} - i\gamma_{n0}$ in Eq \ref{sos_alpha_grnd}, where $\gamma_{n0}$ describes the relaxation mechanisms. The most natural choice of damping parameter is that deriving from Fermi's Golden Rule. That is, the natural linewidth
\begin{align}
\gamma_{nm} &= \frac{1}{\hbar}\Gamma_{nm}\nonumber \\
&= \frac{2}{3\hbar}\left(\frac{\Omega_{nm}}{c}\right)^3e^2|x_{nm}|^2.
\label{natural_linewidth}
\end{align}
This term typically prevents the denominator from vanishing and results in a complex but finite polarizability.  Inclusion of damping, however, has a minimal effect on the divergence seen above because of the form of Eq \ref{natural_linewidth}.  As the system approaches degeneracy, the damping term approaches zero faster than the energy difference, as illustrated in Fig \ref{damping_effects_alpha}.

\subsubsection{Dynamics of an Ensemble}
The dynamics of an ensemble have been treated before \cite{boyd09.01} and are well known. Assuming we are sufficiently close to degeneracy that only the ground and first excited state matter, we can treat the system as consisting of only two levels. The steady-state result for the zero-frequency susceptibility--to all orders--is \cite{boyd09.01}
\begin{align}
\chi = -\frac{N\left(\rho_{11}-\rho_{00}\right)^{(\text{eq})}|\mu_{10}|^2\left(\Omega_{10} + i/T_2\right)T_2^2/\hbar}{1+\left(\omega_{10}\right)^2T_2^2 + (4/\hbar^2)|\mu_{10}|^2|\mathcal{E}|^2T_1T_2}
\end{align}
where $\Omega_{10} = E_{10}/\hbar$ is the resonant frequency between the two states and the relaxation processes are described by time constant $T_1$, the characteristic time at which the population difference $\left(\rho_{11} - \rho_{00}\right)$ decays to its equilibrium value, and $T_2$, the dipole dephasing time between the two states. If we treat the optical field as a reservoir with which the ensemble can exchange energy we can apply the canonical ensemble to compute the probability of being in a given state. The probability $P_i$ of occupying a state with energy $E_i$ is given by the well known expression \cite{landa69.01}
\begin{align}
P_i = \frac{e^{-\beta E_i}}{\sum_j e^{-\beta E_j}},
\label{canonical_probs}
\end{align}
where $\beta = 1/kT$. For a degenerate two-level system, this reduces to
\begin{align}
P_{1,2} = \frac{1}{2}.
\end{align}
which leads to the result
\begin{align}
\left(\rho_{11} - \rho_{00}\right)^{(\text{eq})} = 0.
\end{align}
We see that the steady state response is zero. Thus, the system will only have a large optical response when there is nonzero population difference between the degenerate states; that is, as long as the relaxation processes are very slow compared to the time of the experiment.

\section{Conclusion}
In summary, a fundamental limit has been found on the optical response of excited state systems.  A divergence in the upper limit of the excited state polarizabilities that exceeds the well-known ground state limits motivated the above studies to determine whether this enhancement effect can be applied to real systems. A filtered monte carlo algorithm shows that it is physically feasible for an excited state system to have an intrinsic polarizabilty greater than unity, thus beating the ground state limit. The finite field method demonstrates that the divergent polarizabilities are not simply the result of a misapplication of perturbation theory but are physical in nature on short time scales. Finally, we have shown that the polarizability can be enhanced by exciting a molecule and these enhancements originate in the near degeneracy of highly-excited states that near the continuum band.  Similarly, a system with a degenerate ground state can also show a divergent response; a double quantum well is shown to behave in this way. The relativistic effects on the excited state limits, as well as numerical studies on the effects of degeneracy on the first hyperpolarizability, will be the subject of future studies.

These results suggest several avenues for enhancing the nonlinear-optical response in common systems. In particular, this work suggests the design of new molecules made of two sub-units that are weakly interacting to mimic the particle in a box with a large central barrier.  The difficulty of this approach may be in forcing the system into a ground state where one electron is delocalized over the two sub units. Alternatively, artificial structures such as nanowires with defects or double quantum dots \cite{mi17.01, hayas03.01} may be simpler to fabricate as long as their response is quantum in nature, requiring either low temperature or small structures.  While challenges may persist, these observations demonstrate that there are ways to trick nature into getting ultra large nonlinear response, which can be put to use in new devices.

\appendix
\section{Variation of ansatz}
In this appendix we review the different options for the states to be included in the ansatz. First, we apply the three level ansatz to the excited state quadratic response, however instead of using the states $|0\rangle$, $|1\rangle$, and $|2\rangle$, we use the states $|1\rangle$, $|2\rangle$, and $|3\rangle$. This works out just as the work on ground state $\beta$. We can write the first hyperpolarizability as
\begin{align}
\beta = x_{12}x_{23}x_{31}\left(\frac{2}{E_{21}E_{31}} - \frac{2E_{21} - E_{31}}{E_{31}^3} - \frac{2E_{31} - E_{21}}{E_{21}^3}\right).
\end{align}

From the $(1,1)$ sum rule, we can write
\begin{align}
E_{31}x_{31}^2 \leq E_{21}\left(\tilde{x}_\text{max}^2 - x_{21}^2\right).
\end{align}
Likewise, from the $(2,2)$ sum rule we can write
\begin{align}
E_{32}x_{32}^2 \leq E_{21}\left(\tilde{x}_\text{max}^2 + x_{21}^2\right),
\end{align}
where $\tilde{x}_\text{max}^2 = \frac{\hbar^2}{2mE_{21}}$. Putting this together we can write
\begin{align}
|x_{12}x_{23}x_{31}| &\leq \frac{\tilde{E}}{\sqrt{1 - \tilde{E}}}\tilde{x}_\text{max}^3\tilde{X}\sqrt{1 - \tilde{X}^4}\\
&\equiv \frac{\tilde{E}}{\sqrt{1 - \tilde{E}}}\tilde{x}_\text{max}^3\sqrt[4]{\frac{1}{3}}\sqrt{\frac{2}{3}}G(X),
\end{align}
where $\tilde{E} = E_{21}/E_{31}$, $\tilde{X} = x_{21} / \tilde{x}_\text{max}$, and
\begin{align}
G(X) = \sqrt[4]{3}\sqrt{\frac{3}{2}}\tilde{X}\sqrt{1 - \tilde{X}^4}
\end{align}
has a maximum of unity at $X = 1/\sqrt[4]{3}$. Thus,
\begin{align*}
|x_{12}x_{23}x_{31}| &\leq \frac{\tilde{E}}{\sqrt{1 - \tilde{E}}}\tilde{x}_\text{max}^3\sqrt[4]{\frac{1}{3}}\sqrt{\frac{2}{3}}.
\end{align*}

With the above work, the first hyperpolarizability can thus be written as
\begin{align}
\beta = \sqrt[4]{3}\left(\frac{e\hbar}{\sqrt{m}}\right)^3 \frac{N^{3/2}}{E_{21}^{7/2}} f(\tilde{E}),
\end{align}
where
\begin{align}
f(\tilde{E}) = (1 - \tilde{E})^{3/2} \left(1 + \frac{3}{2}\tilde{E} + \tilde{E}^2\right)
\end{align}
is bounded above by unity. Thus, we have a limit on the first hyperpolarizability for a system for which only the first three excited states are available:
\begin{align}
\beta_\text{max} = \sqrt[4]{3}\left(\frac{e\hbar}{\sqrt{m}}\right)^3 \frac{N^{3/2}}{E_{21}^{7/2}}.
\end{align}
Note that it is simply the ground state limit with $E_{10}\rightarrow E_{21}$, This result should come as no surprise: because we are ignoring the ground state, there can be no spontaneous emission, which is the only mechanism that alters the theory of the fundamental limits. In other words, by leaving the ground state out of the ansatz, the first excited state effectively becomes the ground state.

Here we apply the four-level ansatz.\cite{kuzyk15.01} We include the levels $|0\rangle$, $|1\rangle$, $|2\rangle$, and $|3\rangle$. The expression for $\beta$ becomes
\begin{align}
\beta = \beta_{02} + \beta_{20} + \beta_{03} + \beta_{30} + \beta_{23} + \beta_{32},
\end{align}
where
\begin{align}
\beta_{nm} = -3e^3x_{1n}x_{nm}x_{m1}\left(\frac{1}{E_{n1}E_{m1}} - \frac{2E_{n1} - E_{m1}}{E_{m1}^3}\right) .
\end{align}
The hyperpolarizability is then given by
\begin{align}
\beta = &-3e^3x_{10}x_{02}x_{21}\left(-\frac{2}{E_{10}E_{21}} + \frac{2E_{10} + E_{21}}{E_{21}^3} - \frac{2E_{21} + E_{10}}{E_{10}^3}\right)\nonumber\\
&-3e^3x_{10}x_{03}x_{31}\left(-\frac{2}{E_{10}E_{31}} + \frac{2E_{10} + E_{31}}{E_{31}^3} - \frac{2E_{31} + E_{10}}{E_{10}^3}\right)\nonumber\\
&-3e^3x_{12}x_{23}x_{31}\left(\frac{2}{E_{21}E_{31}} - \frac{2E_{21} - E_{31}}{E_{31}^3} - \frac{2E_{31} - E_{21}}{E_{21}^3}\right) .
\end{align}

We define $\xi_{nm} \equiv x_{nm}/x_\text{max}$, $E = E_{10}/E_{20}$ and $F = E_{10}/E_{30}$ and choose $x = \xi_{10}$, $y=\xi_{20}$, $\xi_{13}$, $E$, and $F$ as free parameters. Using the $(0,0)$, $(1,1)$, and $(2,2)$ sum rules, we can write the transition moments as
\begin{align}
\xi_{02} &= \sqrt{\left(1 - x^2 - y^2/E\right)F}\\
\xi_{12} &= \sqrt{\left(1+x^2 - \frac{1-F}{F}\xi_{13}^2\right)\frac{E}{1-E}}\\
\xi_{23} &= \sqrt{\frac{\left(x^2 - \xi_{13}^2 + 2\right)EF + y^2F + \xi_{13}^2E}{E-F}}.
\end{align}
Similarly, the energy terms can be written in terms of $E$ and $F$. We can then write $\beta$ as
\begin{widetext}
\begin{align}
\beta = \beta_\text{max}^\text{3L} \left(\frac{3}{4}\right)^{3/4}&\left\lbrace xy\left[1+x^2 - \frac{1-F}{F}\xi_{13}^2\right]^{1/2}\sqrt{E}\left[2E(1-E)^2 - (1+E)(1-E)^2E - \frac{2(1-E)^4}{E}\right]\right.\nonumber\\
&+\left.x\xi_{13}\left[1-x^2-\frac{y^2}{E}\right]^{1/2}(1-E)^{7/2}\sqrt{F}\left[\frac{2F}{1-F} - \frac{(1+F)F}{1-F} - \frac{2(1-F)}{F}\right]\right.\nonumber\\
&+\left.\xi_{13}\left[1+x^2 - \frac{1-F}{F}\xi_{13}^2\right]^{1/2}\left[(x^2 - \xi_{13}^2 + 2)EF +y^2F + \xi_{13}^2E\right]^{1/2}\sqrt{\frac{E}{1-F}}\right.\nonumber\\
&\times\left.\left[-\frac{EF(1-E)^2}{1-F}+\left(\frac{1-E}{1-F}\right)^3\frac{2F^3(1-E) - EF^2(1-F)}{E} + \frac{2E^3(1-F) - FE^2(1-E)}{F}\right]\right\rbrace ,
\end{align}
\end{widetext}
which is of the form
\begin{align}
\beta = \beta_\text{max}^\text{3L} \mathcal{L}\left(x, y, \xi_{13}, E, F\right),
\end{align}
where $\beta_\text{max}^\text{3L}$ is the TLA limit and the form of $\mathcal{L}\left(x, y, \xi_{13}, E, F\right)$ can be inferred. Using Mathematica, we can numerically determine $\mathcal{L}_\text{max} = 0.957$, which occurs at the point $(x, y, \xi_{13}, E, F) = (0.011, -0.907, 0.488, 1, 0.450)$ .

It might be surprising that the FLA limit is \textit{smaller} than that of the TLA limit because one might argue that the FLA should reduce to the TLA when all transition moments to/from state $|3\rangle$ vanish, and thus that the FLA limit must be larger than or equal to the TLA limit. Although the FLA has added degrees of freedom, it also has an additional constraint in the form of the $p=q=2$ sum rule. This sum rule was not included to the TLA treatment because it would lead to a contradiction.  An analysis of truncation of the sum rules and which sum rules to use as constraints is a complex one and has been discussed in detail in \cite{kuzyk15.01}, as well as \cite{lytel17.01}. We refer the curious reader to these papers for more information on the subject. However, we can give a simple example of a mathematical system with an objective function and a set of constraints in which an additional degree of freedom reduces the limit of a function, illustrating \textit{how} this can happen.

Consider the function
\begin{align}
\beta = \prod_i x_i
\end{align}
subject to the constraint
\begin{align}
\sum_i x_i = 1 .
\end{align}
For the one-state model we have $x_1 = 1$ and $\beta = x_1 = 1$.  For the two-state model, we have $x_1 + x_2 = 1$, so \[\beta = x_1 x_2 = x_1(1-x_1),\] which is maximum at $x_1 = 1/2$ for a value $\beta = 1/4$.  Note that the added degree of freedom allows $\beta$ to vary but limits its maximum value.

The sum rules are coupled nonlinear equations and the expression for beta is also nonlinear, making it much more difficult to see how the TLA yields a larger limit than the FLA, but the idea is similar: by including an additional degree of freedom (i.e. the state $|3\rangle$), the constraints on the system become more strict (in the form of the $p=q=2$ sum rule), which in turn limits the maximum value of $\beta$.

\bibliography{}

\end{document}